\documentclass[fleqn,usenatbib]{mnras}

\usepackage{newtxtext,newtxmath}

\usepackage[T1]{fontenc}

\DeclareRobustCommand{\VAN}[3]{#2}
\let\VANthebibliography\thebibliography
\def\thebibliography{\DeclareRobustCommand{\VAN}[3]{##3}\VANthebibliography}

\newcommand{\xs}{x^{*}}
\newcommand{\ys}{y^{*}}
\newcommand{\zs}{z^{*}}
\newcommand{\DL}{D_{\rm L}}
\newcommand{\UK}[1]{{\textcolor{black}{#1}}}

\DeclareUnicodeCharacter{2212}{-}
\renewcommand{\arraystretch}{1.2}

\usepackage{graphicx}	
\usepackage{amsmath}	
\usepackage{multirow}

\title{A Bayesian estimator for peculiar velocity correction in cosmological inference from supernovae data}

\author[]{Ujjwal Upadhyay,$^{1,2}$\thanks{E-mail: ujjwalu@iisc.ac.in}
Tarun Deep Saini,$^{1}$
and Shiv K. Sethi$^{2}$
\\
$^{1}$Department of Physics, Indian Institute of Science,
C. V. Raman Road, Bangalore 560012, India \\
$^{2}$Astronomy \& Astrophysics Group, Raman Research Institute,
C. V. Raman Avenue, Bangalore 560080, India
}

\date{Accepted XXX. Received YYY; in original form ZZZ}

\pubyear{\the\year{}}

\begin{document}
\label{firstpage}
\pagerange{\pageref{firstpage}--\pageref{lastpage}}
\maketitle

\begin{abstract}
The peculiar motion of the host galaxies can shift the observed redshifts from their true values, introducing bias in estimating
cosmological parameters from supernova data. The coherent component of peculiar motion is typically corrected using velocity-field reconstruction, while the random component is handled statistically by inflating the magnitude uncertainty through standard error propagation. However, velocity-field reconstruction requires assuming an underlying cosmology, which can bias the final inference, whereas the statistical treatment of the random component relies on a locally linear magnitude–redshift relation and a Gaussian velocity distribution. We present a Bayesian estimator for simultaneously correcting for peculiar motion while fitting a cosmological model to supernova data, relaxing the assumption of linearity of the model and Gaussianity of the peculiar motion. Our approach is based on considering the problem of fitting the magnitude–redshift relation as a non-linear model with errors in both dependent and independent variables. To this end, we develop a general method for fitting non-linear \emph{errors-in-variables} models and apply it to the magnitude–redshift relation, validating it with simulated datasets at the precision of current and upcoming surveys and testing it on the Pantheon sample. Our method provides an alternative approach for accounting for the peculiar velocity effects, which is a complementary method for the coherent component, as it does not require independent velocity measurements, and generalizes the treatment of the random component. Moreover, our general method is applicable to various other problems in cosmology and astronomy.
\end{abstract}




\section{Introduction} \label{sec:first}

The spatially flat $\Lambda$CDM model, founded on the cosmological principle of homogeneity and isotropy, has become the standard description of our observed universe. It has been remarkably successful in describing the observed anisotropies in the Cosmic Microwave Background (CMB)   (\cite{Planck:2018vyg}), large-scale clustering of matter (\cite{DES:2016qvw,2dFGRS:2005yhx,DESI:2024hhd}), and the background expansion history inferred from the redshift and brightness measurements of Type Ia supernovae (SNIa) (\cite{conley2010supernova,Brout:2022vxf,Scolnic:2021amr}).
In addition, a growing number of space- and ground-based missions have ushered in the era of precision cosmology~(\cite{Turner:2022gvw}), which allows us to constrain cosmological parameters at a sub-percent level of precision.
The growing quality and volume of data have enabled us to extract unprecedented information on the structure and evolution of the universe. However, the onset of the era of greater precision has also revealed systematic discrepancies in the standard model of cosmology, such as the observed tensions in the determination of the Hubble constant and the $S_8$ parameter (\cite{DiValentino:2020zio, DiValentino:2020vvd, Abdalla:2022yfr}). Taking these discrepancies as a hint of new physics beyond the standard $\Lambda$CDM model, several modifications to the theory have been proposed in the literature. However, untreated or systematic errors in the analysis as a source of observed discrepancies remain a less explored but plausible alternative. For instance, there are certain anomalies in the CMB observations, such as lensing anomalies, whose presence depends on the method used for statistical analysis. Specifically, the standard {\tt Plik} and {\tt Commander} likelihood of Planck 2018 data prefer a higher value of the lensing parameter $A_L$ than the prediction of the $\Lambda\rm CDM$ model,  $A_L=1$. However, in the likelihood {\tt NPIPE} based on the same Planck data but different modeling of systematics, there is no such anomaly~(\cite{Rosenberg:2022sdy}). It was argued in (\cite{Cortes:2023dij}) that alleviating these tensions with extensions to the $\Lambda$CDM model is not sufficient to claim a hint for physics beyond the  $\Lambda$CDM model. This highlights the fact that increasing the precision in measurement also requires more accurate statistical analysis techniques to obtain consistent results from different observational probes. 

Even with state-of-the-art instrumentation enhanced by artificial intelligence and machine learning for data processing and automation, astronomical observations are never entirely free from errors. These errors can originate from instrumental limitations, atmospheric interference (for ground-based telescopes), or cosmic background noise. Furthermore, systematic biases introduced during data processing and limitations of observational pipelines contribute to these inaccuracies. Despite this, while fitting a model to the data, it is very common to consider errors only on the main observable (dependent variable), treating subsidiary observables (independent variables) as perfectly measured. In part, this is due to the fact that the usual statistical techniques are based on the likelihood function that is derived from the errors in the main observable. This is usually justified if the errors in the subsidiary observables are small compared to the errors in the dependent variable.  In statistics literature, the problem of model fitting when there are errors in both dependent and independent variables is called \emph{errors-in-variables models}~(\cite{dougherty2007introduction}).
Several approaches have been proposed to address this problem when additional information about the independent variable is available~(\cite{dougherty2007introduction}). The \emph{instrumental variable method} introduces an auxiliary variable correlated with the true regressor through a known relation, while the method of \emph{repeated-measurements} reduces uncertainty by averaging multiple observations of the same quantity. In the special case where the model is a monotonic and differentiable function of the regressor, the likelihood can be formulated by inverting the model function ~(\cite{Gregory_2005}). However, there is no general method for fitting arbitrary \emph{errors-in-variable models}. 

In cosmology, this issue could arise on various occasions. Our focus in this work is on fitting magnitude--redshift relation to estimate cosmological parameters from supernova data where peculiar velocities contribute as errors in the redshifts. There are at least two sources of errors in the redshift determination. Spectroscopic redshifts can be determined with a relative accuracy better than $10^{-4}$. The second source of error arises from the line-of-sight peculiar velocities owing to gravitational interaction\footnote{Cosmological peculiar velocities are a key prediction of the gravitational collapse model of structure formation. Their impact can be detected in the redshift-space two-point correlation function of galaxies (e.g. \cite{2020moco.book.....D} and the references therein for details). These velocities can also be detected statistically using the kSZ (Kinetic Sunayaev-Zeldovich) effect (e.g. \cite{Planck:2017xaj,2021A&A...645A.112T})}. The line-of-sight peculiar motion of host galaxies of supernovae contributes to the total observed redshift \footnote{The contribution of peculiar motion also arises from our motion with respect to the cosmic rest frame. The magnitude and direction of this component can be determined by measuring the CMB dipole. It is corrected for in the available SNIa data as in the Pantheon sample.}.  
From linear perturbation theory, the RMS of peculiar velocity varies from a few hundred km/sec to a few thousand km/sec for
scales 1~Mpc to 100~Mpc. The modeling of the peculiar motion of individual objects requires taking into account nonlinear effects, which could further enhance this effect. The impact of peculiar velocities could be more than an order of magnitude larger than the instrumental uncertainty, but still small compared to Hubble recessional velocities except in the nearby universe. 

{ In the supernovae analysis, the large-scale coherent component of peculiar velocity is usually corrected for using velocity field reconstruction based on the observed density distribution~(\cite{Peterson:2021hel,Carreres:2023nmf,Carrick:2015xza,Carr:2021lcj,Giani:2023aor}). This requires assuming an underlying cosmology which can introduce a bias in the cosmological inference. Moreover, this bias will appear differently in fitting different alternatives to the standard model assumed in the reconstruction. The small-scale random component, on the other hand, is treated statistically by propagating the redshift uncertainty due to peculiar motion into the supernova magnitude uncertainty using the standard error propagation method~(\cite{Carreres:2023nmf,Peterson:2021hel}). This involves assuming a locally linear approximation of the magnitude--redshift relation which does not hold at very low redshifts and a Gaussian distribution for the random peculiar motion which can have extended tails at non-linear regime.}

In this work, we present a Bayesian estimator for cosmological parameters which simultaneously corrects for peculiar motion while fitting a cosmological model to the supernova data. It is based on treating each redshift as an independent parameter drawn from a flat prior centered at the position of the measured redshift, which means that no specific underlying cosmological model is assumed for peculiar motion. To this end, we develop a Bayesian method for fitting general non-linear models with errors in both dependent and independent variables, and then specialize it to the case of fitting the magnitude--redshift relation to the supernovae data. For comparison, we analyze the Pantheon SNIa sample (\cite{Pan-STARRS1:2017jku}) using three estimators: one without including peculiar motions ($\mathcal{E}_1$), one based on a locally linear magnitude--redshift relationship, with an additional assumption of Gaussianity of  peculiar velocity ($\mathcal{E}_2$), and finally one that accounts for peculiar motions without making any approximation ($\mathcal{E}_3$).  
To evaluate the efficacy of our estimator, we also applied these estimators to synthetic data with parameters (e.g., number of SNIa, errors on magnitude) typical of the current and future SNIa data sets. For our study, in all cases we consider two cosmological models: a spatially flat $\Lambda$CDM model and a $w$CDM model with constant $w$. 

This paper is organized as follows. In Section~\ref{sec:second}, we discuss in general our method for dealing with the problem of fitting arbitrary models with errors in both variables. In Section~\ref{sec:DZ}, we present the magnitude--redshift relation as an error-in-variables model. In Section~\ref{sec:fourth}, we apply this method to estimate cosmological parameters from supernova data for the $\Lambda$CDM and the $w$CDM model, and discuss the modification in the results as compared to the standard model fitting method that does not consider the contribution of peculiar motion. In Section~\ref{sec:fifth}, we conclude with a discussion of the results and potential future applications of our method. In Appendix \ref{sec:A}, we derive an expression for the likelihood in the linear approximation of the magnitude--redshift relation. 

\section{Methodology} \label{sec:second}
The problem of regression with an arbitrary errors-in-variables model is an interesting problem and, to the best of our knowledge, not a commonly addressed problem in astronomy. Since astronomical data usually have errors in independent variables, this could have general applications in a variety of problems in astronomy. In this section, we describe a Bayesian solution to this problem. In Section~\ref{sec:estimator}, we begin by describing the general background of the problem and develop our estimator by motivating it with the familiar case of independent variables without errors. Then, in Section~\ref{sec:DZ}, we specialize our estimator for estimating cosmological parameters from the magnitude--redshift data of the Pantheon sample of supernovae (\cite{Pan-STARRS1:2017jku}). We consider the contribution of peculiar motion of supernova host galaxies to the redshift uncertainty, which leads to its formulation as an errors-in-variables problem. Later in this paper, we have also compared it to an approximate method that uses a local linear approximation of the magnitude--redshift relation and assumes a Gaussian distribution of errors due to peculiar motions. The development of this method can be found in the Appendix~\ref{sec:A}. This comparison is important because the approximate method is computationally inexpensive and is therefore useful for very large data sets. Our results show that, despite its assumptions, the linear approximation based estimator ($\mathcal{E}_2$) gives comparable results to the exact method. This is because the peculiar velocities are small, and within the variance the magnitude-redshift relationship remains quite close to linear. However, the exact estimator is more flexible in handling a greater variety of problems. 

\subsection{Estimator for Errors-in-Variables}
\label{sec:estimator}

Frequently, experiments produce $N$ pairs $(x_i,y_i)$ of noisy data points with uncertainties (assuming the randomness to be Gaussian) $\sigma_{x_i}$ and $\sigma_{y_i}$ associated with each observed value $x_i$ and $y_i$, respectively.  The quantity $y_i$ often depends on the independently measured quantity $x_i$ through a physical model that is often given in terms of a mathematical relation,
\begin{equation}
      y_{i} = f( \theta,x_i ),
  \end{equation}
or, in general, could be given by an algorithm such as a numerical code or a multistep mathematical operation that outputs $y$ for every $x$. However, our treatment is based on expressing the association of each $x$ with $y$ abstractly through the above relationship.  The function $f(\theta,x_i)$ is generally a function of $m$ unknown model parameters $ \theta \equiv (\theta_1, \theta_2,... \theta_m)$ along with the variable $x_i$.  In the notation used in this paper, if any expression depends explicitly on a single variable, then we notate it with a subscript; else, we use unscripted symbols to refer to them collectively. We shall soon focus on the case of apparent magnitude-redshift data that comprise the measured pair $(z_i, m_{i})$. It might be helpful to the reader to keep this concrete model in mind while reading this section. The parameters $\theta$ would then be the usual cosmological parameters such as the Hubble constant, various matter energy densities, and the equation of state parameter. 

The most straightforward case is where the independent variable $x$ can be measured with negligible error. This is often how such data are fitted to models, often ignoring the $x$ errors even if they are not small. This is due to the fact that the canonical method relies on working with the likelihood function derived from the errors in the dependent variable $y$. Therefore, we must find a way to incorporate $x$ errors as an additional source of error in the likelihood function. If the function $ f( \theta,x_j )$  is a linear function\footnote{Although in general a nonlinear model could refer to a model $ f( \theta,x_j )$ with nonlinear dependence on $\theta$ or $x_i$, in this paper by a nonlinear model we always refer to the latter case,} of $x_j$, then estimating the parameters $ \theta$ from data that have errors in both variables is well established and quite straightforward (see, for example, \cite{Press:1992:NRC} and the derivation in Appendix \ref{sec:A}). This method is based on the conversion of the $x$ error into the $y$ error through the slope of the linear function and adding it in quadrature to the $y$ error while constructing the likelihood function. However, when the model is nonlinear in $x_j$, this trick works only by approximating the fitting function as locally a linear function, as shown for the special case of the magnitude--redshift relation in the Appendix \ref{sec:A}.  Our aim in this paper is to develop a method that is not approximate or ad hoc and treats errors in $x$ and $y$ on equal footing.  This turns out to be possible in Bayesian statistics, as we show next.

\subsubsection{Estimator for nonlinear models } \label{sec:nonlinear}

We introduce our approach through the familiar case of data with errors only in the dependent variable $y$ and none in the independent variable $x$. Let the true values of $y$ be denoted by $\ys$. The observed values of $y$ differ from their true value by a random number, $y_i= \ys_i +\epsilon_{Yi}$, where $\epsilon_{Yi}$ is the random error in measurement $y_i$. Since $x$ are measured without error, $\ys_i = f( \theta,x_i )$.  If $P_{\rm Y} (\epsilon_Y)$ is the probability distribution function for random noise in the measurement of $y$, the likelihood function is given by
\begin{equation}
     \mathcal{L}_Y( \mathcal{D}\mid \theta, \ys) = \prod_i  P_{Y}( y_i- \ys_i)\delta(\ys_i -f( \theta,x_i )),
   \label{eq:nonlin0}  
\end{equation}
where $\mathcal{D}$ collectively refers to the pair $(x,y)$. In this form, the likelihood function can only be evaluated if we know the values of $\theta$ and $\ys$. The Dirac delta function ensures that the likelihood is non-zero only when $\ys_i = f( \theta,x_i )$. Since $x_i$ is known but not $\ys_i$, this equation can be satisfied for any value of $\ys_i$ at $\theta$ differing from their true value. Furthermore, the subscript $Y$ in the likelihood function has been added to make it explicit that it is derived from the random noise in $y$.  In a sense, we can think of the true values $\ys$ as additional parameters of the model, which is how they are treated in the following analysis. 

Using Bayes' theorem, this likelihood function can be used to write the posterior probability distribution for the parameters $\theta $, and $\ys$ as  
\begin{equation}
    P(\theta, \ys \mid \mathcal{D};I) \propto  \mathcal{L}_Y( \mathcal{D}\mid \theta, \ys) P(\theta \mid I) 
\end{equation}
where $P(\theta \mid I)$ is the prior probability distribution function for $\theta$,  and the addition of $I$ in various terms symbolizes all the prior information that we may have (for details, see, e.g., \cite{Sivia1996}). We also assumed that there is no prior information on the true values $\ys$. Combining Eq.~(\ref{eq:nonlin0}) with this equation  we obtain 
    
\begin{equation}
    P(\theta, \ys \mid \mathcal{D};I) \propto
  \prod_i  P_{Y}( y_i- \ys_i)\delta(\ys_i -f( \theta,x_i ))  P(\theta \mid I)\,. 
\end{equation}

This posterior probability distribution has information about the true values $\ys$, which is of little use. Usually, the parameters $\theta$ are of interest; therefore, we can marginalize over $\ys$ by integrating over them to obtain the most commonly used expression for the posterior probability for a model with errors only in dependent variable, 
\begin{equation}
    P(\theta \mid \mathcal{D};I) \propto  \prod_i  P_{Y}( y_i- f( \theta,x_i )) P(\theta \mid I) \,.
\end{equation}

To generalize this to the case with errors in both variables, in analogy with $y$ we write $x_i = \xs_i + \epsilon_{Xi}$,  where $\xs$ is the unknown true value of $x$ in terms of which the likelihood function can be written as 
\begin{equation}
     \mathcal{L}_{XY}( \mathcal{D}\mid \theta, \xs,\ys) = \prod_i  P_{Y}( y_i- \ys_i) P_{X}( x_i- \xs_i)\delta(\ys_i -f( \theta,\xs_i )),
   \label{eq:jointlikelihood}  
\end{equation}
  where we have used the fact that $x$ and $y$ are statistically independent, therefore, their joint probability can be obtained simply by multiplying their individual probability distribution functions. However, note that since $x$ are now noisy, the argument of delta function now contains only true  values $(\xs, \ys)$. The modified posterior distribution is now given by
\begin{equation}
      P(\theta, \xs,\ys \mid  \mathcal{D};I) \propto \prod_i  P_{Y}( y_i- \ys_i) P_{X}( x_i- \xs_i)\delta(h(\xs,\ys;\theta) )P(\theta\mid I)
      \label{eq:sympost}
\end{equation}
where we have defined $h(\xs,\ys;\theta) \equiv \ys_i -f( \theta,\xs_i) = 0$
as a symmetric way to write the model relationship between $x$ and $y$.  The obvious advantage is that it makes the difference between dependent and independent variables disappear. It is worth noting that in general, the data could have several independent variables $(x_i,y_i,z_i,\cdots)$ related through $h(x_i,y_i,z_i,\cdots)=0$. It is straightforward to generalize this analysis to such multivariate cases.  

The posterior distribution in Eq.~(\ref{eq:sympost}) is a function not only of the model parameters but also of the unknown true values of $x$ and $y$ for each data point. If we are interested only in the model parameters $\theta$, then we can integrate over additional parameters. This yields
\begin{equation}
      P(\theta \mid  \mathcal{D};I) \propto \int d^N\xs\prod_i P_{Y}( y_i- f(\theta, \xs_i)) P_{X}( x_i- \xs_i)P(\theta\mid I)
\end{equation}

If we assume Gaussian errors in both variables with standard deviations $\sigma_y$ and $\sigma_x$, respectively, then after defining 
\begin{equation}
   \chi^2_Y(\theta, \xs) = 
\sum_{i=1}^{N} \left[\frac{y_i-f( \theta,\xs_i)}{\sigma_{y_i}}\right]^2, \quad
\chi^2_X(\xs)  = \sum_{i=1}^{N} \left[\frac{\xs_i-x_i}{\sigma_{x_i}}\right]^2\,, 
  \end{equation}
the marginalized posterior distribution can be written as
\begin{equation}
    P(\theta \mid \mathcal{D}; I) \propto  \int d^N\xs \exp \left[-\frac{\chi^2_Y (\theta, \xs)+ \chi^2_X(\xs)}{2}  \right] P(\theta\mid I)\,.
\end{equation}
This form is used in Appendix~\ref{sec:A} to derive an approximate form of the estimator where integration over $\xs$ can be done analytically. However, if we do not assume a specific form for $P_{X}$ then assuming Gaussian errors only in $y$ we get 
\begin{equation}
    P(\theta \mid \mathcal{D}; I) \propto  \int d^N\xs \exp \left[-\frac{\chi^2_Y (\theta, \xs)}{2} \right] P_{X}( x_i- \xs_i) P(\theta\mid I)\,.
    \label{eq:finfor}
\end{equation}

In practice,  the integration in Eq.~(\ref{eq:finfor}) is more conveniently carried out with  Markov Chain Monte Carlo (MCMC), as we discuss later. Since the total number of model parameters now exceeds the number of data points, there is, in principle, the possibility of overfitting the data. In order to obviate this possibility, we add a regularization term to ensure that the physical model satisfy the condition that the mean value of $\chi_Y^2$  is close to the degrees of freedom of the model\footnote{This expectation is technically true only for linear models, but holds well in general for small errors as models can then be roughly thought of as locally linear.}. This choice of regularization is motivated by the fact that, under Gaussian measurement errors, a well-specified model is expected to yield residuals that follow a $\chi^2$-distribution with a mean equal to the number of degrees of freedom. Enforcing this condition would constrain the effective volume of the latent-parameter space without introducing additional model-dependent assumptions, thereby suppressing unphysical solutions while preserving the physical content of the likelihood. The regularisation is therefore expected to be effective in preventing overfitting in situations where the model is sufficiently smooth, and the noise in the dependent variable is well described by Gaussian statistics. However, a precise understanding of when such regularisation is needed and how effective it is in practice remains to be established and requires further investigation, which is a topic of future work. The marginalized $\xs$ can be considered as subsidiary parameters not directly related to the physical model. Therefore, the degrees of freedom are given by the number of data points, $N$, minus the number of physical parameters, $m$. For the case of Gaussian noise in the dependent variable, the required regularization term is just the probability distribution function for $\chi_Y^2$ given by
\begin{equation}
     P_{\chi^2} (\chi^2_Y) = \frac{\chi_Y^{k-2} \exp(-\chi^2_Y/2)}{2^{k/2}\Gamma(k/2)},
\end{equation}
where $k=N-m$ is the desired degrees of freedom of the fit. For the case of fitting magnitude-redshift relationship for distant supernovae, this would be the total number of supernovae minus the number of cosmological parameters being fitted.  Therefore, our final form for the posterior distribution function for $\theta$ can be obtained by multiplying it with the probability distribution function for $\chi_Y^2$ 
\begin{equation}
\label{eq:nonlin2}
    P(\theta \mid \mathcal{D}; I) \propto  \int d^N\xs \exp \left[-\frac{\chi^2_Y ( \theta, \xs)}{2}  \right] P_{X}( x_i- \xs_i)P({\theta}\mid I) P_{\chi^2} (\chi_Y^2)\,.
\end{equation}

Regularization is commonly used to fit complex models with a large number of parameters as it acts as a constraint and facilitates stable maximization of the posterior distribution (e.g. \cite{Sivia1996}).  However, as we note later, the addition or omission of this term may not always change the dimensionality of the fit, as is further explained in the context of the specific cosmological application of this method presented in the paper. It is clear that the method advocated here is more expensive to implement numerically since each independent variable is a parameter. We chose cosmological data from SNIa as a proof of concept for our proposed method.

Further, integrating the RHS of the above equation over the model parameters, we get the expression for the Bayesian evidence,
\begin{equation}
   {\mathcal{Z} = \int d\theta \int d^N\xs \exp \left[-\frac{\chi^2_Y ( \theta, \xs)}{2}  \right] P_{X}( x_i- \xs_i)P({\theta}\mid I) P_{\chi^2} (\chi_Y^2)\,.}
\end{equation}
Since marginalisation and regularisation can significantly affect the Bayesian evidence, particularly when the number of latent parameters, $\xs$, is large, the application of our estimator to evidence-based model comparison is non-trivial and should be treated with caution. In particular, for fully flexible models, the interplay between model flexibility, prior volume, and regularisation can strongly influence the evidence, and the present analysis does not establish that the method is generically safe for Bayesian model selection. Evidence-based model comparison in such settings, therefore, remains an open question and requires further study before the approach can be applied reliably.
\section{Magnitude--Redshift Relation} \label{sec:DZ}
Although the measurement error in the observed redshift of a supernova host galaxy is negligibly small, due to the possible peculiar motion of the host galaxy relative to a comoving observer, its observed redshift is shifted with respect to the theoretical cosmological redshift that encodes the expansion history of the universe.  To estimate this shift, we note that 
\begin{equation}
    \frac{\nu_{0}}{\nu_{\rm obs}} = \frac{\nu_{0}}{\nu_{\rm com}} \frac{\nu_{\rm com}}{\nu_{\rm obs}} \approx (1+{v_p}/{c})(1+z)
\end{equation}
where $\nu_{\rm obs}, \nu_0$, and $\nu_{\rm com}$ are the supernova photon frequencies observed on earth, in the rest frame of the galaxy, and by a comoving observer at the position of the galaxy; and $v_{p}$ is the component of velocity of a galaxy with respect to the comoving observer, parallel to the line of sight from the observer to the supernova host galaxy.  

Peculiar motions also cause beaming of light towards the observer, alter the observed flux, and cause additional k-correction (e.g. see \cite{1984MNRAS.206..377E} for the combined impact of all these effects on the number count). The first two effects act in concert, as an object moving towards an observer would appear closer and less redshifted.  The standard analysis already takes into account measurement errors in the luminosity of an SNIa, and these errors are much larger than what we expect from the peculiar motion of host galaxies.   However, this is not true for redshift measurement errors that, as already noted, are small compared to peculiar motions. At low redshifts, this effect is enhanced as peculiar redshift/blueshift can become comparable to the redshift of the host galaxy. As more and more supernovae are observed, the effective statistical error in luminosity distance reduces at any redshift because of statistical averaging.  Thus, we need techniques that allow us to handle errors on redshift, even if they are small. The impact of these peculiar velocities on the inference of cosmological parameters has been studied using the standard error propagation method~(\cite{2011ApJ...741...67D}) and N-body simulations~(\cite{Wojtak:2013gda,Wu:2017fpr,Peterson:2024mjx,PhysRevD.111.043516}). In the standard supernova analysis, the effect of coherent component of peculiar velocity is taken into account using velocity field reconstruction based on the observed galaxy distribution, while the uncertainty due to the random component is treated statistically by propagating it to the magnitude uncertainty using a locally-linear approximation to the magnitude--redshift relation and assuming Gaussian distribution for the random motion (See Appendix \ref{sec:A}). Our Bayesian estimator simultaneously accounts for both coherent and random motions, without the need for independent velocity measurements or the simplifying approximations mentioned above.

The luminosity distance versus redshift relation is usually expressed in terms of the apparent magnitude of distant supernovae. To apply the analysis given in the previous section, we set the redshift $z_i$ equal to $x_i$ and the apparent magnitude $m_i$ to the function $f(\theta,x_i)$. The apparent magnitude $m$ and the luminosity distance $\DL$ of a supernova of known absolute magnitude $M$ are related through
\begin{equation}
\label{eq:lumdist}
  m=M+  5\log\left(\frac{\DL}{\text{Mpc}}\right) +25\;.  
\end{equation}
For a flat Friedmann–Lemaître–Robertson–Walker universe, the luminosity distance of a source at redshift $z$ can be expressed as
\begin{equation}
    \DL = \frac{c(1+z)}{H_0}\int_0^z\frac{dz'}{h(z')} ,
\end{equation}
where the dimensionless Hubble parameter is given by
\begin{equation}
    h(z) = \left [\Omega_{\rm M}(1+z)^3 +\Omega_{\rm DE}(1+z)^{3(1+w)}\right ]^{1/2}\,,
\end{equation}
where $\Omega_{\rm M}$ and $\Omega_{\rm DE}$ are the present-day matter and dark energy densities, respectively. The dark energy and pressure are related through the parameter $w$ through $p_{\rm DE} = w\rho_{\rm DE}$. If dark energy is assumed to be the cosmological constant, then $w=-1$ in this parametrization. The apparent magnitude can be written as
\begin{eqnarray}
 m =M &+& 5\log(cH^{-1}_0/{\rm Mpc}) \nonumber\\
 &+&5 \log \left[(1+z)\int_0^z\frac{dz'}{h(z)}\right] + 25\,.
\end{eqnarray}
Here, $w$, $\Omega_{\rm M}$, and $H_0$ are free parameters of the model. However, note that $H_0$ is degenerate with $M$, which allows us to estimate only their combination. 

Specializing the method described in the previous section to the magnitude–redshift relation, we obtain our estimator for the cosmological parameters that simultaneously account for the unknown peculiar velocities.
Here, we present the three \emph{maximum a posteriori} (MAP) estimators used in this work in standard notation for ease of reference and comparison:
\begin{equation}
    \mathcal{E}_k: \hat{\theta}_{\mathrm{MAP}} = \arg\max_{\theta} \, \Big[P_k(\theta \mid \mathcal{D}; I)\Big], k=1,2,3.
\end{equation}
where
\begin{equation}
     P_1(\theta \mid \mathcal{D}; I) \propto \prod_i^N \exp \left[-\frac{1}{2}\frac{\left(m_i-m(z_i)\right)^2}{\sigma^2_{m_i} }\right] 
\end{equation}

\begin{equation}
     P_2(\theta \mid \mathcal{D}; I) \propto \prod_i^N \exp \left[-\frac{1}{2}\frac{\left(m_i-m(z_i)\right)^2}{\sigma^2_{m_i} + m'^2(z_i)\; \sigma^2_{z_i}}\right] 
\end{equation}

\begin{eqnarray}
     P_3(\theta \mid \mathcal{D}; I) \propto  \prod_i^N \int d^N\zs_i \exp \left[-\frac{1}{2}\frac{(m_i-m(\zs_i))^2}{\sigma^2_{m_i} }\right] \;\;\; \nonumber \\
     P_{Z}( z_i- \zs_i) P({\theta}\mid I) P_{\chi^2} (\chi_m^2)\,.
\end{eqnarray}

\begin{itemize}
    \item The first estimator, labeled $\mathcal{E}_1$, neglects redshift uncertainties and assumes that the redshifts are perfectly measured. It serves as a baseline case used for comparison with the other estimators to highlight the impact of redshift errors and peculiar velocities on cosmological parameter estimation.  

    \item The second estimator,  $\mathcal{E}_2$, corresponds to the standard approach to accounting for random peculiar motions in cosmological parameter estimation. It relies on the linearized magnitude--redshift relation that incorporates redshift uncertainties through the likelihood function given in Eq.~(\ref{eq:linapprox}) and derived in Appendix~\ref{sec:A}. This estimator further assumes that the redshift errors follow a Gaussian distribution.  

    \item The final MCMC-based estimator, $\mathcal{E}_3$, employs Eq.~(\ref{eq:nonlin2}) described in Section~\ref{sec:nonlinear}. It explicitly accounts for the contribution of peculiar velocities to the observed redshift by treating each redshift as a uniformly distributed random variable within a range around its measured value. This generalizes $\mathcal{E}_2$ by incorporating model nonlinearity and arbitrary redshift error distributions. Moreover, it is fundamentally different from $\mathcal{E}_2$, enabling a self-consistent treatment of coherent peculiar motions as well.
\end{itemize}

Although the estimator $\mathcal{E}_2$ could be derived as a limiting case of the estimator $\mathcal{E}_3$, under the locally-linear approximation of the magnitude--redshift relation and the assumption of Gaussian distribution for the redshift errors (See Appendix \ref{sec:A}), $\mathcal{E}_3$ is conceptually very different from $\mathcal{E}_2$. While $\mathcal{E}_2$ treats the redshift error by propagating it to the magnitude and thereby effectively de-weighting the low redshift SNIa in the posterior, $\mathcal{E}_3$ treats the redshifts as parameters and looks for maximizing the likelihood in the space of redshift parameters. Thus, any shift in cosmological parameters estimated using the standard estimator $\mathcal{E}_2$ compared to completely neglecting the random peculiar velocities (as in $\mathcal{E}_1$) is due to a change in relative weights of low- and high-redshift SNIa, while a similar shift in the $\mathcal{E}_3$ estimate is due to moving the observed redshift towards the true cosmological redshift. 

\section{Results}\label{sec:fourth}
In cosmological studies, our primary concern is with the parameters $w$ and $\Omega_{\rm M}$. Although distant SNIa are not ideal for determining $H_0$  due to degeneracy with $M$, we also present results for the Hubble parameter for the $\Lambda$CDM and the $w$CDM models. To break the degeneracy, we use the absolute magnitude $M=-19.24$  corresponding to the SH0ES~(\cite{Riess:2021jrx}) value of  $H_0=73.04$ km/s/Mpc for the Pantheon sample.

Note that only the line-of-sight component of the peculiar velocity contributes to the change in redshift. By the isotropy of the peculiar velocity dispersion, we have $\sigma_{v_{\rm 3D}}=  \sqrt{3}\sigma_{v_{p}}$. Thus, any assumption we make about the magnitude of peculiar velocity must keep this in mind. For estimator $\mathcal{E}_3$, the distribution $P_Z$ is minimally assumed to be uniform in a specified redshift range. More specifically, we assume that each observed $z_i$ in the data deviates from its true value $z^*_i$, which falls within a specified uncertainty range $z^*_i \in [z_{i}^{\rm min},z_{i}^{\rm max}]$, where the min and max values are chosen to correspond to our assumptions about peculiar velocities. We also need to assign a value to $\sigma_{z_i}$ in Eq.~(\ref{eq:linapprox}). We can do that by setting the range chosen for $\mathcal{E}_3$ to be three times $\sigma_{z_i}$. This ensures that the one-sigma range from $-\sigma_{z_i}$ to $\sigma_{z_i}$ for a Gaussian distribution containing 99.7\% probability is equal to the redshift range allowed in $\mathcal{E}_3$. Therefore, $\sigma_{z_i} = (z_{i}^{\rm max}-z_{i}^{\rm min})/6$. Using $\Delta z = ({v_p/c})(1+z)$,   the standard deviation used in $\mathcal{E}_2$ is given by $\sigma_z = (\sigma_{v_p}/c)(1+z)$. Throughout, we use $\sigma_{v_p} = 250\;\mathrm{km/s}$. For the estimator $\mathcal{E}_3$, a uniform prior $[-1000, 1000] \; \rm km/s$ has been used for the peculiar velocity $v_p$. Further increasing the prior range mainly worsens the sampling efficiency and requires longer convergence times, without providing additional benefit. Using a more restrictive prior, on the other hand, reduces the freedom of the latent redshift parameters and leads the posterior obtained from the estimator $\mathcal{E}_3$ closer to that from $\mathcal{E}_1$, as expected. We initialize the redshift parameters at their observed values, $z_i$, which is a convenient and natural choice, and find that as long as the prior range is reasonably large, the sampling converges to the true redshifts, $z_i^*$. Thus, the cosmological posteriors behave as expected and are robust to reasonable variations in the prior. We also observe that the redshift parameters are well constrained by the data at low redshifts, where they significantly affect the inference of cosmological parameters, while at higher redshifts they are largely prior-dominated, as the likelihood is not sensitive to them and does not influence the inference.

\begin{figure*}
    \centering
    \begin{minipage}[b]{0.48\textwidth}
        \centering
        \includegraphics[scale=0.7]{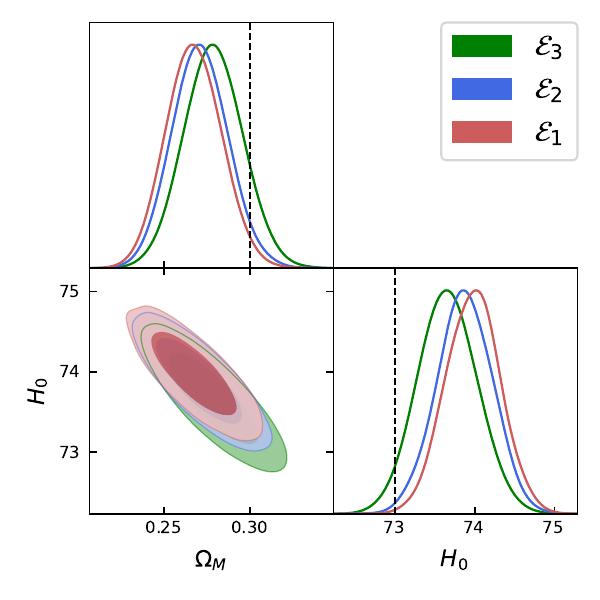}
        \vspace{-0.5em}
    \end{minipage}
    \hfill
    \begin{minipage}[b]{0.48\textwidth}
        \centering
        \includegraphics[scale=0.7]{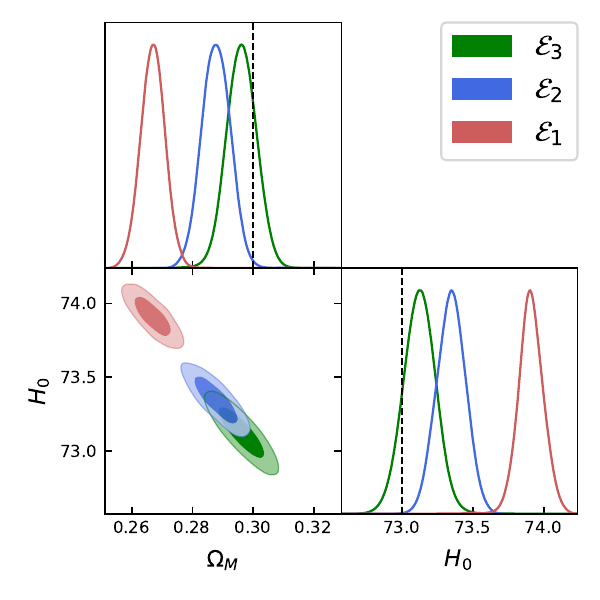}
        \vspace{-0.5em}
    \end{minipage}

   \caption{Constraints on the $\Lambda$CDM parameters from synthetic data with  magnitude error  $\sigma_m = 0.2$ (left) and $\sigma_m = 0.05 $ (right). Estimator $\mathcal{E}_1 $ (red) ignores any contribution to redshift from the peculiar motion of supernovae host galaxies, $\mathcal{E}_2$ (blue) considers it for the linearized model (see Appendix~\ref{sec:A} for details), while $\mathcal{E}_3$ (green) represents our estimator, which incorporates the contribution from peculiar motion for the exact model. The black dashed line represents the true values of the parameters.  The errors on the estimated parameters are marginally larger for estimators $\mathcal{E}_2$ and  $\mathcal{E}_3$.}
    
    \label{fig:LCDM_combined_syn}
\end{figure*}

\renewcommand{\arraystretch}{1.5}
\setlength{\tabcolsep}{4pt}
\begin{table*}
    \centering
    \footnotesize
    \begin{tabular}{|c c c c|c c c|}
        \hline
        \multirow{2}{*}{\textbf{Model}} & \multirow{2}{*}{\textbf{Parameter}} & \multirow{2}{*}{\textbf{Prior}} & \multirow{2}{*}{\textbf{True value}} & \multicolumn{3}{c|}{\textbf{Best-fit value $\pm$ 1$\sigma$ uncertainty}} \\
        \cline{5-7}
        & & & & ${\mathcal{E}_1}$ & ${\mathcal{E}_2}$ & ${\mathcal{E}_3}$ \\
        \hline
        \multirow{2}{*}{$\Lambda$CDM} 
        & $\Omega_{\rm M}$ & $\mathcal{U}(0.1, 0.5)$ & $0.30$ & $0.267\pm 0.016$ & $0.271\pm 0.017$ & $0.279\pm 0.017$ \\
        & $H_0$ [km/s/Mpc] & $\mathcal{U}(60, 80)$ & $73.0$ & $73.97\pm 0.35$ & $73.87\pm 0.36$ & $73.66\pm 0.38$ \\
        \hline
        \multirow{3}{*}{$w$CDM} 
        & $\Omega_{\rm M}$ & $\mathcal{U}(0.1, 0.5)$ & $0.30$ & $0.360^{+0.041}_{-0.028}$ & $0.357^{+0.041}_{-0.031}$ & $0.327^{+0.060}_{-0.041}$ \\
        & $H_0$ [km/s/Mpc] & $\mathcal{U}(60, 80)$ & $73.0$ & $74.97\pm 0.71$ & $74.84\pm 0.73$ & $73.93\pm 0.78$ \\
        & $w$ & $\mathcal{U}(-3, 0)$ & $-1.0$ & $-1.43^{+0.21}_{-0.19}$ & $-1.40^{+0.22}_{-0.18}$ & $-1.19^{+0.24}_{-0.19}$ \\
        \hline
    \end{tabular}
    \caption{Priors and best-fit values for the $\Lambda$CDM and $w$CDM parameters obtained from synthetic data with $\sigma_m = 0.2$. Results are shown for three estimators: $\mathcal{E}_1$, which ignores peculiar velocities; $\mathcal{E}_2$, which uses the linear approximation; and $\mathcal{E}_3$, which models them exactly.}
    \label{tab:lcdmsyn}
\end{table*}

For presenting our results, we need to marginalize over various parameters, which means integrating over them in the posterior distribution.  Also note that in Eq.~(\ref{eq:nonlin2}), we need to integrate over the variable $\xs$, which corresponds to the redshift $z^*$ in this case.  Clearly, it is not feasible to numerically integrate so many variables. Therefore, to carry out marginalization, we employ the MCMC sampling technique. Specifically, we implement the Metropolis-Hastings algorithm within the MCMC framework to draw samples from the posterior distribution and use GetDist~(\cite{Lewis:2019xzd}) Python library for the analysis of MCMC samples. For all cases, we run 12 independent MCMC chains for $10^5$ steps. We use the Gelman-Rubin criterion of convergence. For estimators $\mathcal{E}_1$ and $\mathcal{E}_2$, we achieve convergence with $R - 1 < 0.01$, while for $\mathcal{E}_3$ we obtain $R - 1 < 0.05$. It might be surprising that we are able to find the optimal solution even with more than a thousand parameters. To achieve this, we need to start our MCMC runs from the observed values of redshifts. However, as long as the redshifts are assumed to have peculiar velocity components not exceeding 1000 km/s, the starting point does not matter too much. The extension of this interval reduces the acceptance rate of MCMC and increases $ R-1$ for a fixed number of steps. Thus, it leads to slower convergence. Since the cosmological line-of-sight peculiar velocities are expected to be much smaller than 1000 km/s, we have not investigated a larger range. 

We found that MCMC sampling requires a longer time to converge if we treat cosmological parameters and redshift parameters on par. However, the efficiency of convergence is significantly improved by drawing multiple samples of redshifts for the same cosmological parameters. Thus, in all the results below, MCMC samples the redshift five times more often than the cosmological parameters. This ensures that MCMC is able to optimally sample the parameter space. \UK{We also find that our results are not altered significantly by regularization (Eq.~(\ref{eq:nonlin2})). The regularization term only changes the acceptance rate and marginally changes the final $\chi^2$ per degree of freedom, indicating that the inclusion of redshift errors does not increase the dimensionality of our fitted physical model. This is not surprising if we note that overfitting requires the fitting function to be able to pass through most of the noisy data points that are displaced vertically due to errors in magnitude. This would require the fitting function to have the flexibility that Eq.~(\ref{eq:lumdist}) does not have with respect to treating the redshift as a parameter. The function remains monotonic, ensuring that the convergence is robust even without regularization.} However, we emphasise that while the estimator performs well in the specific case of fitting the magnitude--redshift relation, the present work does not establish its general applicability. In particular, understanding the behaviour of the estimator in more general settings, especially the role and effectiveness of regularisation in fully flexible dark-energy models, remains an open issue that we leave for future work. The present study should therefore be viewed as a proof of principle that requires further investigation, rather than as a fully developed method.

\subsection{Application to Synthetic Data}
\label{sec:B}

\begin{figure*}
    \centering
    \begin{minipage}[t]{0.48\textwidth}
        \centering
        \includegraphics[scale=0.45]{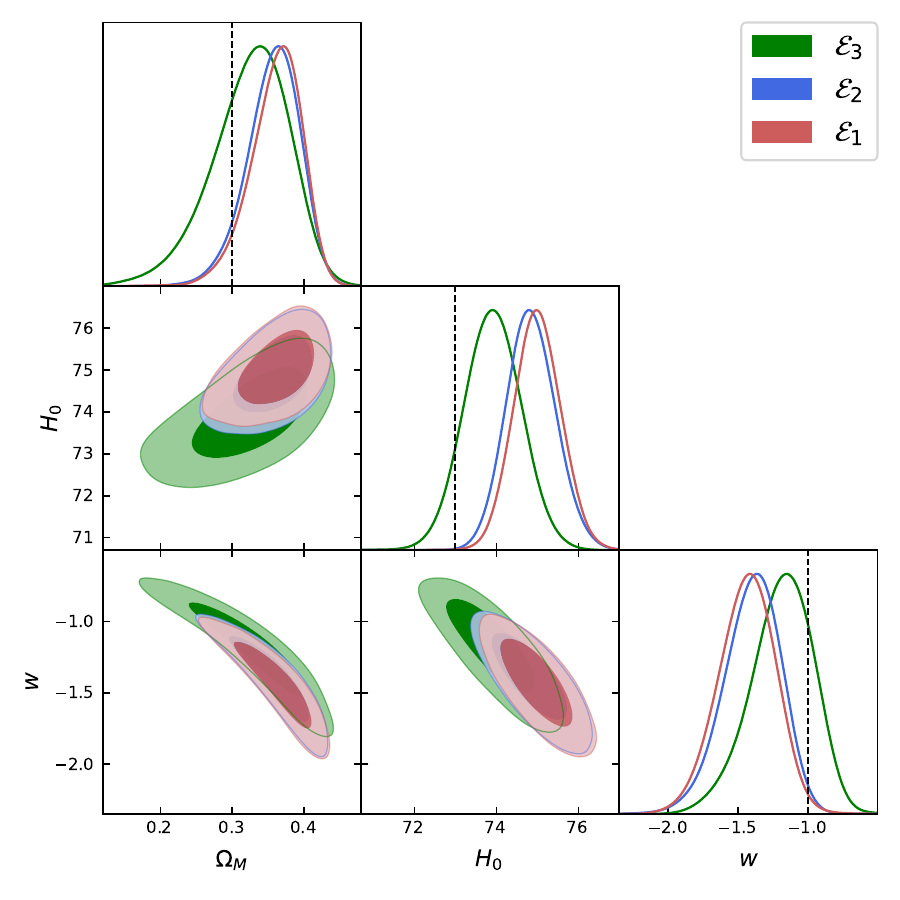}
        \vspace{-0.5em}
    \end{minipage}
    \hfill
    \begin{minipage}[t]{0.48\textwidth}
        \centering
        \includegraphics[scale=0.45]{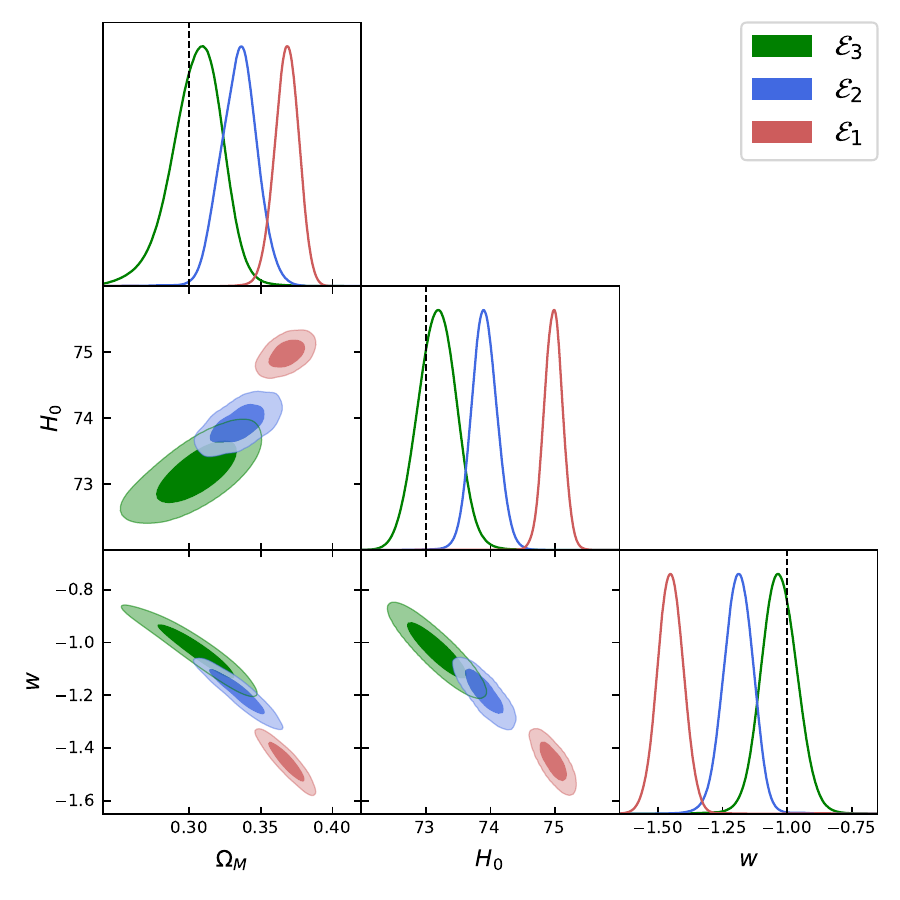}
        \vspace{-0.5em}
    \end{minipage}

    \caption{Constraints on the $w$CDM parameters from synthetic data with  $\sigma_m = 0.2$ (left) and $\sigma_m = 0.05$ (right). The colour scheme of Figure~\ref{fig:LCDM_combined_syn} is followed. The black dashed line represents the true values of the parameters.
    As in Figure~\ref{fig:LCDM_combined_syn}, the errors on the estimated parameters are marginally larger for estimators $\mathcal{E}_2$ and  $\mathcal{E}_3$. In addition, the analysis of synthetic data shows that the neglect of peculiar velocities might bias the estimation of cosmological parameters. The  Figure in conjunction with  Figure~\ref{fig:LCDM_combined_syn} shows that this bias might not play an important role in the current data (left panel), but it would become important in the interpretation of the future data (right panel).  }
    \label{fig:wcdm_combined}
\end{figure*}

\renewcommand{\arraystretch}{1.5}
\setlength{\tabcolsep}{4pt}
\begin{table*}
    \centering
    \footnotesize
    \begin{tabular}{|c c c c|c c c|}
        \hline
        \multirow{2}{*}{\textbf{Model}} & \multirow{2}{*}{\textbf{Parameter}} & \multirow{2}{*}{\textbf{Prior}} & \multirow{2}{*}{\textbf{True value}} & \multicolumn{3}{c|}{\textbf{Best-fit value $\pm$ 1$\sigma$ uncertainty}} \\
        \cline{5-7}
        & & & & ${\mathcal{E}_1}$ & ${\mathcal{E}_2}$ & ${\mathcal{E}_3}$ \\
        \hline
        \multirow{2}{*}{$\Lambda$CDM} 
        & $\Omega_{\rm M}$ & $\mathcal{U}(0.1, 0.5)$ & $0.30$ & $0.2670\pm 0.0051$ & $0.2879\pm 0.0050$ & $0.2962\pm 0.0050$ \\
        & $H_0$ [km/s/Mpc] & $\mathcal{U}(60, 80)$ & $73.0$ & $73.91\pm 0.11$ & $73.35\pm 0.12$ & $73.12\pm 0.12$ \\
        \hline
        \multirow{3}{*}{$w$CDM} 
        & $\Omega_{\rm M}$ & $\mathcal{U}(0.1, 0.5)$ & $0.30$ & $0.3675^{+0.0093}_{-0.0076}$ & $0.335\pm 0.014$ & $0.305^{+0.019}_{-0.015}$ \\
        & $H_0$ [km/s/Mpc] & $\mathcal{U}(60, 80)$ & $73.0$ & $74.96^{+0.15}_{-0.13}$ & $73.89^{+0.21}_{-0.18}$ & $73.17^{+0.31}_{-0.28}$ \\
        & $w$ & $\mathcal{U}(-3, 0)$ & $-1.0$ & $-1.449\pm 0.067$ & $-1.189\pm 0.062$ & $-1.031\pm 0.072$ \\
        \hline
    \end{tabular}
    \caption{Priors and best-fit values for the $\Lambda$CDM and $w$CDM parameters obtained from synthetic data with $\sigma_m = 0.05$. Results are shown for three estimators: $\mathcal{E}_1$, which neglects peculiar velocities; $\mathcal{E}_2$, which uses a linear approximation; and $\mathcal{E}_3$, which models them exactly.}
    \label{tab:wcdmsyn}
\end{table*}

 We evaluate the efficacy of our estimators by testing them on synthetic data. In order to enable a better comparison with the real observations, we adopt the parameters of the Pantheon sample comprising 1048 supernovae and their redshifts range $z\in[0.01,2.26]$, in our simulations. As mentioned briefly in the previous section, our method generalizes the standard approach to accounting for the random component of peculiar motion by incorporating the full non-linearity of the model and relaxing the assumption of Gaussianity for the peculiar velocity distribution. More importantly, our approach is fundamentally different from the standard one, as the standard method merely propagates redshift uncertainties to the inferred cosmological parameters, whereas our framework also shifts the observed redshifts toward their true values, thereby correcting for the effects of coherent peculiar velocities. To show this, we simulate data with both random and coherent components of peculiar velocity. For the random component, we add redshift errors drawn from a Gaussian distribution with zero mean and standard deviation corresponding to peculiar velocity, $v_p=250 \;\rm km/s$, while for the coherent component, we introduce a constant shift to the redshifts of nearby ($z<0.03$) SNe~Ia as a proxy for large-scale bulk motion of $300 \;\rm km/s$. Although a constant shift does not represent a true coherent velocity field, it is sufficient  to underline the advantage of  the proposed estimator  over  the standard one. 
 
We analyze two cosmological models, $\Lambda$CDM and $w$CDM, considering two different cases of supernova magnitude uncertainties. For the first case, we fixed the quality of the data at roughly the same level as Pantheon data, assuming the supernova magnitudes to have Gaussian errors with mean zero and standard deviation $\sigma_m = 0.2$. 
 For the second case, we consider a future SNIa survey with smaller magnitude errors $\sigma_m = 0.05$.  Such small errors are possible with a sample 16 times larger\footnote{This data quality might be achieved with the Zwicky Transient Facility (ZTF) \cite{Graham:2019qsw} and the Large Synoptic Sky Survey Telescope (LSST) \cite{LSSTScience:2009jmu, LSST:2008ijt}, albeit with higher supernova density at $z < 0.5$, see, e.g.\cite{2020PDU....2900519G} for details).} than the Pantheon sample with the magnitude errors obtained by binning the data in redshift. In order to quantify the effect of reduced errors on the measured magnitude, we used the same redshifts as  in the first case. Moreover, the random errors used to generate the data are also the same, except for a reduction of a factor of four to achieve $\sigma_m = 0.05$. The cosmological and other parameters are fixed to the following values: $\Omega_{\rm M}=0.30$, $H_0=73 ~\rm km~s^{-1}~Mpc^{-1}$, $w=-1.0$, and the absolute magnitude of the supernova $M=-19.24$.
 
 Our main results are shown in Figures~\ref{fig:LCDM_combined_syn} and~\ref{fig:wcdm_combined} together with Tables~\ref{tab:lcdmsyn} and~\ref{tab:wcdmsyn}. We carried out many more simulations for our study  and the results presented here are representative of the observed trends. A comparison of our analysis of synthetic data with the input cosmology enables us to discern the following:

 1) Even with the current precision in supernova magnitudes (Figure~\ref{fig:LCDM_combined_syn}, left panel) and the expected peculiar velocities assumed in the simulation, neglect of  peculiar velocity effects ($\mathcal{E}_1$) can significantly bias the inferred cosmological parameters. The standard estimator, $\mathcal{E}_2$, accounts only for the random component, leading to larger parameter uncertainties and only a slight shift of the mean values toward the true inputs (black dashed). This small shift is expected and primarily arises from the reduced weighting of low-$z$ SNe~Ia data  rather than an explicit peculiar-velocity correction. In contrast, the proposed estimator, $\mathcal{E}_3$, provides estimates that are statistically more consistent with the true values, even though the overall differences among the estimators remain small. 

 2) The impact of peculiar velocities becomes more pronounced with improved precision in supernova magnitudes (Figure~\ref{fig:LCDM_combined_syn}, right). A similar trend is observed, but with a more statistical significance: while the estimator $\mathcal{E}_2$ shifts the inferred parameters toward the true values, it remains inconsistent with them. In contrast, $\mathcal{E}_3$ recovers the true values within less than $1\sigma$. Interestingly, the estimates from $\mathcal{E}_2$, although not fully consistent with the true values, appear closer than expected, given that this estimator is not designed to correct for peculiar-velocity effects but merely to propagate the associated uncertainties. Again, this behavior arises from the de-weighting of low-$z$ SNe~Ia in the posterior. The effect of peculiar velocities is strongest at low redshifts, where their contribution becomes comparable to the cosmological redshift. Since $\mathcal{E}_1$ neglects this effect and treats the observed redshift as purely cosmological, the model fits the data poorly, leading to a biased estimate. In contrast, $\mathcal{E}_2$ incorporates redshift uncertainties, which propagate into magnitude uncertainties (see Eq.~21). These uncertainties are larger at low redshift, where the slope of the magnitude–redshift relation is steeper, effectively down-weighting the low-$z$ SNe~Ia. As a result, the cosmological constraints are dominated by high-$z$ supernovae, where peculiar-velocity effects are smaller, and the estimated parameters lie closer to the true values. Moreover, the effect of this de-weighting becomes more pronounced as the supernova magnitude uncertainty decreases, since the redshift uncertainty propagated into the magnitude then begins to dominate in the total error budget.

3) In the analysis of the $w$CDM model shown in Figure~\ref{fig:wcdm_combined}, we observe a trend similar to that in the $\Lambda$CDM case (Figure~\ref{fig:LCDM_combined_syn}). The proposed estimator, $\mathcal{E}_3$, accounts for both the random and coherent components of peculiar velocities, yielding an unbiased estimate of the cosmological parameters. In contrast, the standard estimator considers only the random component by propagating redshift uncertainties into the magnitude.

 \subsubsection{Coherent vs Random Peculiar Motion}
To examine how our method performs for the random and coherent components individually, in this section, we analyze the $\Lambda$CDM model by treating these components separately. Figures~4 and~5 show the posterior distributions of the $\Lambda$CDM parameters obtained from synthetic data simulated with redshift uncertainties arising solely from the random and coherent components of peculiar motion, respectively. The levels of peculiar velocity used are the same as in the previous case, with the random component having zero mean and a standard deviation of $250~\mathrm{km,s^{-1}}$, and the coherent component set to $300~\mathrm{km,s^{-1}}$. The supernova magnitude uncertainty is fixed at $\sigma_m = 0.05$. 

 \begin{figure}
    \centering
     \includegraphics[scale=0.75]{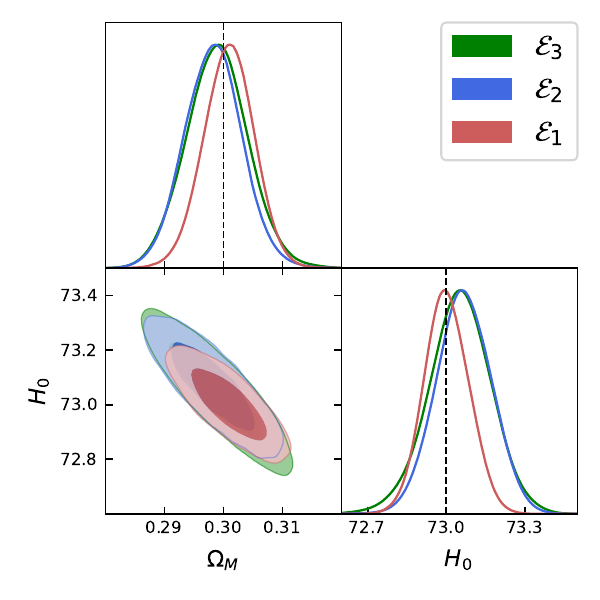} 
        \caption{Comparison of different estimators with only a random peculiar motion. The colour scheme of Figure~\ref{fig:LCDM_combined_syn} is followed. Both $\mathcal{E}_2$ and $\mathcal{E}_3$ give similar results, and the standard estimator $\mathcal{E}_2$ is sufficient.}
        \label{fig:lcdm_rng}
\end{figure}

 \begin{figure}
    \centering
     \includegraphics[scale=0.75]{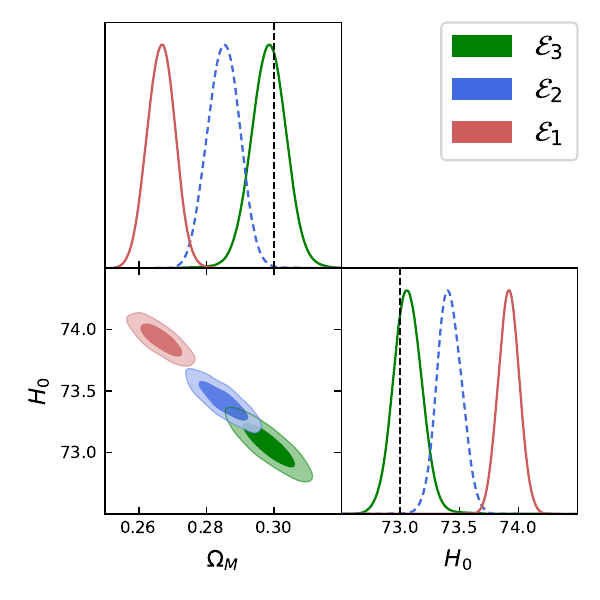} 
        \caption{Comparison of different estimators with only the coherent component of peculiar motion. The colour scheme of Figure~\ref{fig:LCDM_combined_syn} is followed. As expected, $\mathcal{E}_2$ fails to capture the coherent component of peculiar motion, as it is constructed to handle the random contribution. It is therefore not directly comparable with $\mathcal{E}_3$ and is displayed with a dashed line. The estimator $\mathcal{E}_3$, however, correctly accounts for the coherent component. }
        \label{fig:lcdm_rng}
\end{figure}

In Figure~4, both the standard estimator $\mathcal{E}_2$ and our estimator $\mathcal{E}_3$ yield nearly identical results, with slightly larger uncertainties compared to $\mathcal{E}_1$, as expected. This is because both $\mathcal{E}_2$ and $\mathcal{E}_3$ propagate the redshift uncertainty arising from random peculiar motions into the cosmological parameters, thereby increasing the uncertainty in the estimates. The estimator $\mathcal{E}_3$ is a generalization of $\mathcal{E}_2$, incorporating model nonlinearity and non-Gaussian redshift errors; however, in this case, it naturally converges to $\mathcal{E}_2$ since, for peculiar velocities of $\sim250~\mathrm{km,s^{-1}}$ and the lowest redshifts ($z \sim 0.01$), the magnitude–redshift relation remains well approximated by a linear form. Moreover, the simulated random motions are Gaussian. The small shift observed in the estimated parameters is statistically insignificant. In Figure~5, we find that the estimator $\mathcal{E}_3$ effectively corrects for the coherent component, yielding cosmological parameters consistent with the true input values, with only a marginal increase in the associated uncertainties. This correction arises because the redshifts are treated as free parameters that can vary around their observed values, allowing them to converge toward the true redshifts during sampling. Once the redshift parameters reach their stationary states, small random oscillations persist around the best-fit values, with amplitudes determined by the precision of the supernova magnitudes. As the supernova data become more precise, this residual uncertainty is expected to diminish. The result from the standard estimator $\mathcal{E}_2$ is shown in dashed lines as it is not a proper comparison, since $\mathcal{E}_2$ is not constructed for correcting the coherent component. The apparent shift of the $\mathcal{E}_2$ estimates toward the true values is instead purely a consequence of de-weighting, which effectively suppresses the contribution of low-redshift supernovae. 

\subsection{Application to Pantheon Data}
\begin{figure}
    \centering
     \includegraphics[scale=0.75]{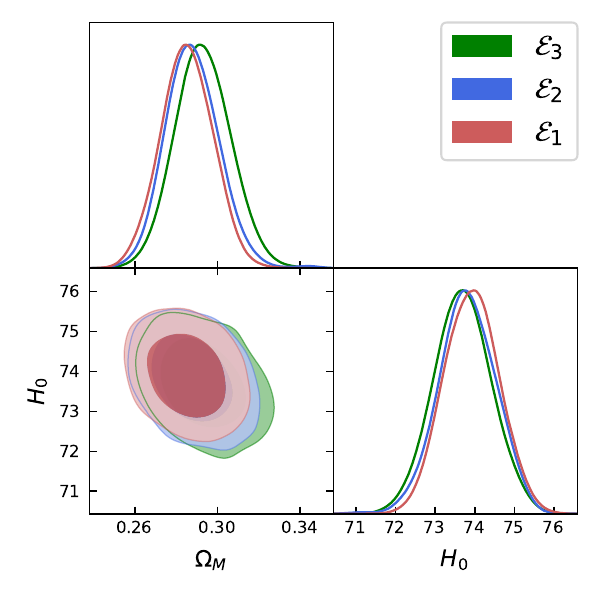} 
       \caption{The figure shows posterior distributions and credible regions of  $\Lambda$CDM parameters from Pantheon data using three different estimators. It follows the convention used in Figure~\ref{fig:LCDM_combined_syn}. All the estimators give statistically consistent results.    \\ }
        \label{fig:lamcdm}
\end{figure}

\begin{figure}
    \centering
    \includegraphics[scale=0.50]{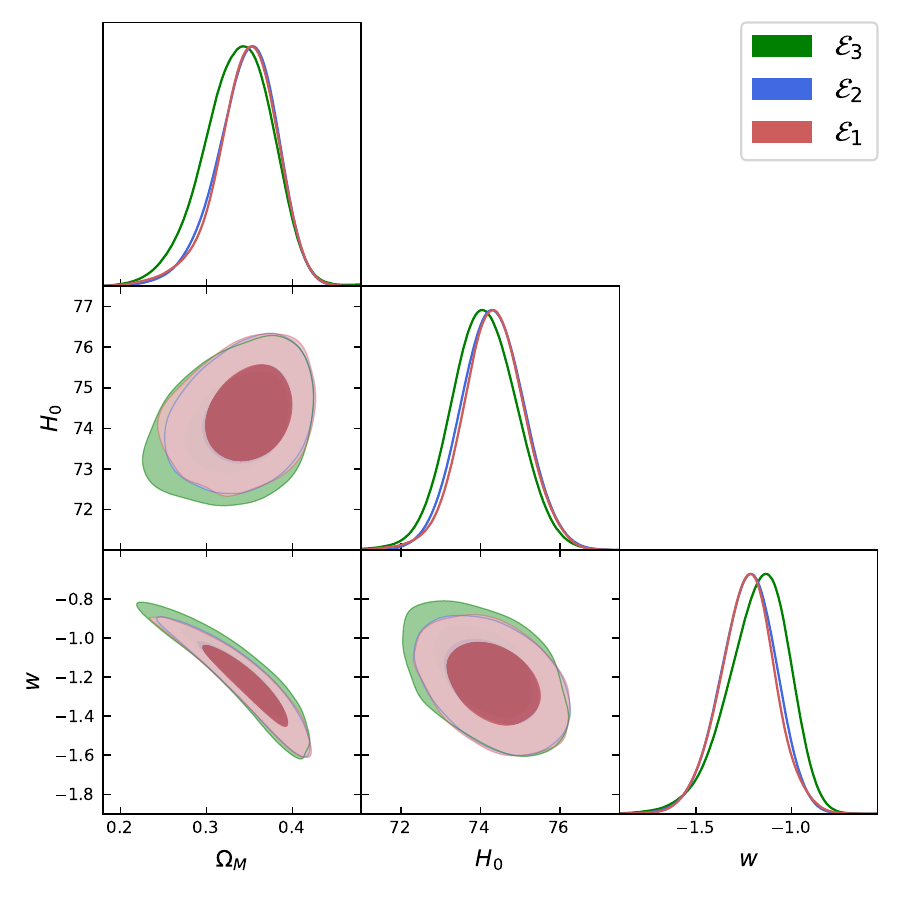} 
  \caption{The figure shows posterior distributions and credible regions of  $w$CDM parameters from Pantheon data using three different estimators. It follows the convention used in Figure~\ref{fig:LCDM_combined_syn}. As in Figure~\ref{fig:lamcdm}, all the estimators give statistically consistent results. However, the observed small shift in the mean values obtained from $\mathcal{E}_2$ and $\mathcal{E}_3$ could indicate a residual coherent component of peculiar motion in the data.\\ }
  \label{fig:wcdm}
\end{figure}

We also applied our method to fit both the $\Lambda$CDM and $w$CDM models to the Pantheon sample of SNe~Ia, using all three estimators for comparison and the same color scheme as in the previous analyses. The posterior distributions with the corresponding $1\sigma$ and $2\sigma$ credible regions are shown in Figures~\ref{fig:lamcdm} and~\ref{fig:wcdm}, and the numerical results are summarized in Table~\ref{tab:Pan}. In both cosmological models, we adopt uniform priors for the parameters, as listed in the corresponding tables.

In Figures~\ref{fig:lamcdm} and~\ref{fig:wcdm}, both the baseline estimator $\mathcal{E}_1$ and the standard estimator $\mathcal{E}_2$ yield nearly identical results, while the estimate from $\mathcal{E}_3$ shows a slight shift, consistent with the trend observed in Figures~\ref{fig:LCDM_combined_syn} and~\ref{fig:wcdm_combined}. However, owing to the relatively large uncertainty in the supernova magnitudes, all three estimates remain statistically consistent, indicating that the standard estimator is adequate for cosmological parameter estimation in this case. This is expected, since the coherent velocity component has already been corrected for in the Pantheon data. Nevertheless, a small shift between $\mathcal{E}_2$ and $\mathcal{E}_3$ could indicate the presence of a residual coherent component or a non-Gaussianity in the random peculiar velocities, as $\mathcal{E}_2$ and $\mathcal{E}_3$ completely agree for purely Gaussian random motions (see Figure~4). Additionally, in Figure~\ref{fig:wcdm_combined}, a larger shift is observed between the cosmological parameters estimated using $\mathcal{E}_2$ and $\mathcal{E}_3$. In particular, the value of $w$ obtained from $\mathcal{E}_3$ is more consistent with $\Lambda$CDM, suggesting that even with the current level of precision, a small residual systematic in the redshift could mimic an apparent evolution in dark energy if not properly corrected.

In summary, although the results from the Pantheon data do not change significantly when using our generalized estimator—which accounts for model nonlinearity, non-Gaussian random peculiar motions, and the coherent velocity component—the implications are nonetheless important. Even a small deviation in the estimates obtained with our method could point to residual peculiar motions or unaccounted systematics that may become evident with improved observational precision. In particular, analyzing SNe~Ia data without properly accounting for peculiar velocities can bias the inferred nature of dark energy, a key issue in both cosmology and theoretical physics that has gained renewed attention following the recent DESI results (e.g., \cite{DESI:2025zgx}).
 

\renewcommand{\arraystretch}{1.5}
\setlength{\tabcolsep}{4pt}
\begin{table*}
    \centering
    \footnotesize
    \begin{tabular}{|c c c c|c c c|}
        \hline
        \multirow{2}{*}{\textbf{Model}} & \multirow{2}{*}{\textbf{Parameter}} & \multirow{2}{*}{\textbf{Prior}} & \multirow{2}{*}{\textbf{True value}} & \multicolumn{3}{c|}{\textbf{Best-fit value $\pm$ 1$\sigma$ uncertainty}} \\
        \cline{5-7}
        & & & & ${\mathcal{E}_1}$ & ${\mathcal{E}_2}$ & ${\mathcal{E}_3}$ \\
        \hline
        \multirow{2}{*}{$\Lambda$CDM} 
        & $\Omega_{\rm M}$ & $\mathcal{U}(0.1, 0.5)$ & $0.30$ & $0.285 \pm 0.013$ & $0.288^{+0.012}_{-0.014}$ & $0.293\pm0.014$ \\
        & $H_0$ [km/s/Mpc] & $\mathcal{U}(60, 80)$ & $73.0$ & $73.90 \pm 0.70$ & $73.80 \pm 0.74$ & $73.69\pm0.73$ \\
        \hline
        \multirow{3}{*}{$w$CDM} 
        & $\Omega_{\rm M}$ & $\mathcal{U}(0.1, 0.5)$ & $0.30$ & $0.347^{+0.038}_{-0.028}$ & $0.346^{+0.039}_{-0.029}$ & $0.337^{+0.043}_{-0.036}$ \\
        & $H_0$ [km/s/Mpc] & $\mathcal{U}(60, 80)$ & $73.0$ & $74.31\pm 0.85 $ & $74.30\pm 0.80 $ & $74.03^{+0.90}_{-0.77}$ \\
        & $w$ & $\mathcal{U}(-3, 0)$ & $-1.0$ & $-1.23\pm 0.15$ & $-1.23^{+0.15}_{-0.14}$ & $-1.18^{+0.18}_{-0.14}$ \\
        \hline
    \end{tabular}
    \caption{Priors and best-fit values for the $\Lambda$CDM and $w$CDM parameters obtained from Pantheon data. Results are shown for three estimators: $\mathcal{E}_1$, which neglects peculiar velocities; $\mathcal{E}_2$, which uses a linear approximation; and $\mathcal{E}_3$, which models them exactly.}
    \label{tab:Pan}
\end{table*}

\section{Conclusion and Discussion}
\label{sec:fifth}
Cosmology has become a precision science in the last few decades. This has been achieved by an unprecedented increase in the quality and quantity of data, namely, Planck CMB temperature and polarization maps, large galaxy surveys such as SDSS, the detection of thousands of SNIa, etc. This also means that physical effects hitherto considered beyond the ability of astronomical instruments are well within reach. However, such rapid progress also requires one to continually improve existing statistical techniques and develop new ones. This paper is an effort in that direction. 

In this paper, we have developed a general Bayesian estimator (described in Section~\ref{sec:nonlinear}) for fitting non-linear models to data with uncertainties in both independent and dependent variables, and applied it to two cosmological models using both synthetic and observational data to assess its efficacy and relevance for current and future SNe~Ia surveys. Our analysis demonstrates that the estimator effectively accounts for both the random and coherent components of peculiar velocities of SNe~Ia host galaxies in cosmological parameter estimation. When applied to synthetic data with uncertainties comparable to current observations, we find that the impact of peculiar velocities remains small due to the relatively large magnitude errors, making the standard estimator—accounting only for random motions—sufficient. A similar behavior is observed in the Pantheon data, supporting the validity of our assumptions regarding peculiar velocities. However, future surveys with higher precision in the supernovae magnitude, such as the Zwicky Transient Facility (ZTF;~\cite{Graham:2019qsw}) and the Legacy Survey of Space and Time (LSST;~\cite{LSSTScience:2009jmu, LSST:2008ijt}), will be more sensitive to peculiar velocity effects. While the larger supernova samples will reduce the influence of random motions through statistical averaging, the contribution from coherent flows will become increasingly significant, necessitating more accurate methods to correct for it. The statistical framework presented in this work thus provides a robust alternative to existing approaches for extracting cosmological parameters from these forthcoming data sets.

Our approach to account for the effects of peculiar motions in cosmological inference from SNe~Ia data is fundamentally different from previous studies that rely on velocity-field reconstruction for the coherent component and the standard linear-approximation estimator for the random component. The standard estimator assumes a locally linear magnitude–redshift relation and a Gaussian distribution for the random peculiar velocities, thereby propagating the redshift uncertainty into the cosmological parameters. Our method generalizes this framework by relaxing these assumptions. Although the linear and Gaussian approximations are valid for current SNe~Ia data—and will likely remain so for future surveys—they may break down with the inclusion of very low-redshift supernovae, where linearity no longer holds. In such cases, our approach offers a more accurate treatment. The principal advantage of our method lies in its treatment of the coherent component: unlike velocity-field reconstruction, which relies on galaxy density measurements and assumes a fiducial cosmology for the reconstruction (potentially biasing subsequent cosmological inference), our approach corrects for coherent motions self-consistently during model fitting by treating redshifts as free parameters alongside the cosmological ones. Consequently, our framework simultaneously accounts for both random and coherent peculiar velocities while estimating the cosmological parameters, avoiding model-dependent biases inherent in reconstruction-based methods. Furthermore, the general estimator developed here for non-linear \emph{errors-in-variables} models extends naturally to other inference problems in cosmology and astronomy.

Although our method can correctly estimate cosmological parameters in fairly general settings, including cases with correlated and non-Gaussian peculiar-velocity fields, it is computationally expensive. In addition to the number of SNe, the computational cost depends on the sampling strategy and on the degree of correlation among the redshift parameters. In our analysis, we find that the cost scales approximately linearly with the number of SNe when the redshift parameters are uncorrelated. For future large samples, such as those expected from LSST, a more detailed investigation of computational optimisation and scaling behavior will be necessary, which we leave for future work.

In future work, we also aim to extend our analysis with a better modeling of peculiar velocities and explore the feasibility of detecting them directly from SNe~Ia data, which could provide valuable insights into the growth of structure and the nature of gravity on cosmological scales. Furthermore, a combined analysis of multiple SNe~Ia data sets—such as Pantheon$+$, DESY5, and Union3—may enhance the precision of our results, particularly in constraining the nature of dark energy, which we plan to investigate in subsequent studies. We also anticipate that the general framework developed in this work will be applicable to the analysis of other cosmological data sets, which we intend to explore in future research.

\section*{Acknowledgments} \label{sec:acknowledgments}
UU would like to acknowledge Yashi Tiwari for her valuable input throughout the project, particularly her insightful feedback on improving the efficiency of our Python code. UU also thanks Somnath Bharadwaj, Purba Mukherjee, and Avinash Paladi for their helpful comments. We acknowledge the use of the HPC cluster of the Raman Research Institute, Bangalore, India.

\section*{Data Availability}

The Pantheon sample of Type Ia supernovae data analyzed in this work is publicly available.


\bibliographystyle{mnras}
\bibliography{references} 


\appendix

\section{Linear Approximation}
\label{sec:A}
The problem of fitting errors-in-variable models is quite straightforward for linear models. This is because the linear model allows us to marginalize the likelihood over the true regressor analytically, resulting in a very simple likelihood that depends only on the observed values of the variables~(\cite{Press:1992:NRC}). Since our model is not linear, we cannot apply it exactly. We compare our method with the standard method applied to the linear approximation of the model. Here, we derive an expression for the likelihood of the linear approximation of our model.\\
\\Apparent magnitude as a function of redshift, $z$, is given by
\begin{equation}
  m(z)=M+5\log\left(\frac{\DL(z)}{\text{Mpc}}\right) +25\;,  
\end{equation}
where $\DL(z)$ is the luminosity distance given by,
\begin{equation}
    \DL(z)=(1+z)\int_0^z\frac{c\;dz}{H(z)
} .
\end{equation}
Considering errors in the measurement of both the apparent magnitude, $m$, and redshift, $z$, and assuming the errors to be Gaussian distributed, the posterior can be expressed as

\begin{equation}
     P(\theta \mid \mathcal{D}; I) \propto \prod_i^N \exp \left[ -\frac{1}{2}\left(\frac{m_i-m(z^{*}_i)}{\sigma_{m_{i}}}\right)^2 -\frac{1}{2}\left(\frac{z_i-z^{*}_i}{\sigma_{z_i}}\right)^2 \right]
\end{equation}

This expression cannot be used to constrain the parameters of the model from the data since we do not have the true redshift values $z_*$, only the observed values $z$. Taylor expanding the apparent magnitude about the observed redshift and keeping terms up to linear order in redshift, we can write
\begin{equation}
    m(z^*)=m(z) + m'(z) (z^*-z)
    \label{eq:Taylor}
\end{equation}
where $m'(z)$ is the derivative of $m(z^*)$ w.r.t. $z^*$ evaluated at $z^*=z$, and is given by

\begin{equation}
    m'(z)=\frac{5}{\ln 10}\left[\frac{1}{1+z}+\frac{1}{H(z)\int_0^z\frac{dz^*}{H(z^*)}}\right]
\end{equation}
\\With the above approximation, Eq.~(\ref{eq:Taylor}), the posterior takes the form,

\begin{eqnarray}
&&P(\theta \mid \mathcal{D}; I) \propto 
\prod_{i=1}^N 
\exp \Bigg[ 
-\frac{1}{2} \left( \frac{m_i - m(z_i)}{\sigma_{m_i}} \right)^2  
-\frac{1}{2} \Bigg\{ 
\left( \frac{m'^2(z_i)}{\sigma^2_{m_i}} + \frac{1}{\sigma^2_{z_i}} \right)\times \nonumber \\
&&\quad (z_i^* - z_i)^2 
- \frac{m'(z_i)\, [m_i - m(z_i)]}{\sigma^2_{m_i}} (z_i^* - z_i) 
\Bigg\} 
\Bigg].
\end{eqnarray}

Integrating over the true but unknown redshift $z^*$ using the standard Gaussian integral, we obtain the marginalized posterior,
\begin{equation}
    P(\theta \mid \mathcal{D}; I) \propto \prod_i^N \exp \left [-\frac{1}{2}\frac{\left(m_i-m(z_i)\right)^2}{\sigma^2_{m_i} + m'^2(z_i)\; \sigma^2_{z_i}} \right]
     \label{eq:linapprox}
\end{equation}

Now, the right-hand side of the posterior depends only on the observed data. If we do not know the values of $\sigma_{z_i}$, as in our case, we can fix it at some value or make it a parameter. In this work, we chose to fix it at $\sigma_z= (\sigma_{v_p}/c)(1+z)\;$, with $\sigma_{v_p} = 250$ km/s. Effectively, this choice sets a broad prior on the redshifts or the peculiar velocities.


\bsp	
\label{lastpage}

\end{document}